\documentclass[sigconf, nonacm]{acmart}

\usepackage[linesnumbered,ruled,vlined]{algorithm2e}
\usepackage{listings}
\usepackage{float} 
\usepackage{tikz}
\usepackage{pgfplots}
\usetikzlibrary{positioning}
\usetikzlibrary{patterns}

\usepackage{xcolor}

\definecolor{codegray}{gray}{0.9}
\definecolor{keywordcolor}{rgb}{0.26, 0.38, 0.96}
\definecolor{stringcolor}{rgb}{0.1, 0.5, 0.1}
\definecolor{commentcolor}{rgb}{0.5, 0.5, 0.5}

\lstdefinestyle{pythonstyle}{
    backgroundcolor=\color{white},
    language=Python,
    basicstyle=\ttfamily\small,
    keywordstyle=\color{keywordcolor}\bfseries,
    stringstyle=\color{stringcolor},
    commentstyle=\color{commentcolor}\itshape,
    showstringspaces=false,
    breaklines=true,
    frame=single,
    tabsize=4
}

\lstdefinestyle{mystyle}{
    breaklines=true,        
    basicstyle=\ttfamily\footnotesize,  
    frame=single,           
    keepspaces=true,        
    columns=flexible,       
    captionpos=b            
}

\SetCommentSty{mycommfont}

\SetKwInput{KwInput}{Input}                
\SetKwInput{KwOutput}{Output}

\newcommand\vldbyear{2026}
\newcommand\vldbworkshop{AIDB}
\newcommand\vldbauthors{\authors}
\newcommand\vldbtitle{\shorttitle} 
\newcommand\vldbavailabilityurl{https://github.com/gsvic/Jailbreak}
\newcommand\vldbpagestyle{plain}

\begin{document}
\title{Breaking Database Lock-in: Agentic Regeneration of High Performance Storage Readers for Database Bypass}

\author{Victor Giannakouris}
\affiliation{%
  \institution{Cornell University}
  \city{Ithaca, NY}
  \country{USA}
}
\email{vg292@cornell.edu}

\author{Immanuel Trummer}
\affiliation{%
  \institution{Cornell University}
  \city{Ithaca, NY}
  \country{USA}
}
\email{itrummer@cornell.edu}


\begin{abstract}
Analytical workloads operating on data stored in external database systems face a fundamental bottleneck: data access is guarded entirely by the database driver, like JDBC or ODBC, forcing all reads through query execution and other driver layers that are not designed for bulk columnar analytics. Existing approaches, like JDBC and ODBC connectors, inherit this overhead, resulting in significant latency and throughput limitations for downstream external, analytic query engines. We present Jailbreak, an approach that bypasses the database engine entirely by reading storage files directly and materializing data as in-memory columnar buffers. Jailbreak's key insight is that database file formats, while complex, are fully specified by their source code and documentation, artifacts that Large Language Models (LLMs) can ingest to regenerate operator-specific table reading components without human-engineered parsing logic. 
Jailbreak leverages LLM-assisted code synthesis for database storage decoding, turning a traditionally opaque format into a directly queryable artifact.
We evaluate Jailbreak on PostgreSQL and MySQL storage files, targeting analytical snapshot scenarios common in read replicas and offline processing pipelines. The generated reader produces Apache Arrow buffers consumable directly by most of the widely known query engines, including DuckDB, Apache Spark, and GPU-accelerated frameworks such as cuDF and Spark RAPIDS. We validate correctness against JDBC/ODBC-based baselines using the TPC-H benchmark across all query results, and demonstrate significant performance improvements in end-to-end analytical throughput, achieving up to 27$\times$ speedups. Our results suggest that LLM-assisted storage reader synthesis is a viable and generalizable methodology for breaking data lock-in across database systems, with applications beyond PostgreSQL and MySQL for any system whose file format is available to the LLM from documentation or source code.

\end{abstract}

\maketitle

\pagestyle{\vldbpagestyle}
\begingroup\small\noindent\raggedright\textbf{VLDB Workshop Reference Format:}\\
\vldbauthors. \vldbtitle. VLDB \vldbyear\ Workshop: \vldbworkshop.\\ 
\endgroup
\begingroup
\renewcommand\thefootnote{}\footnote{\noindent
This work is licensed under the Creative Commons BY-NC-ND 4.0 International License. Visit \url{https://creativecommons.org/licenses/by-nc-nd/4.0/} to view a copy of this license. For any use beyond those covered by this license, obtain permission by emailing \href{mailto:info@vldb.org}{info@vldb.org}. Copyright is held by the owner/author(s). Publication rights licensed to the VLDB Endowment. \\
\raggedright Proceedings of the VLDB Endowment. 
ISSN 2150-8097. \\
}\addtocounter{footnote}{-1}\endgroup

\ifdefempty{\vldbavailabilityurl}{}{
\vspace{.3cm}
\begingroup\small\noindent\raggedright\textbf{VLDB Workshop Artifact Availability:}\\
The source code, data, and/or other artifacts have been made available at \url{\vldbavailabilityurl}.
\endgroup
}

\section{Introduction}
\label{sec:introduction}
The boundary between relational database systems (RDBMSes) and analytical query engines has long been guarded by the RDBMS engine itself.
When an external system or application, like a GPU-accelerated framework, a vectorized query engine, or an ML feature store, needs data that resides in PostgreSQL or MySQL, it must issue SQL queries via a \emph{wire protocol}, like JDBC or ODBC, wait for the server to traverse its storage layer, serialize rows through the wire protocol, and then deserialize those rows on the client side before any useful computation can begin.
This architecture made sense when databases were monolithic systems that owned both storage and compute, but it is increasingly at odds with the modern data stack, in which a growing ecosystem of specialized engines, like DuckDB~\cite{raasveldt2019duckdb}, DataFusion~\cite{datafusion}, cuDF~\cite{cudf}, Spark SQL~\cite{zaharia2016apache} or Velox~\cite{pedreira2022velox}, expects data in columnar, zero-copy formats such as Apache Arrow~\cite{arrow}.
 
The overhead of this engine-mediated path is well documented.
ConnectorX~\cite {wang2022connectorx} showed that more than 85\% of the wall-clock time in a typical \texttt{read\_sql} call is consumed by client-side deserialization and format conversion, not by query execution or network transfer.
Raasveldt and M\"{u}hleisen~\cite{raasveldt2017don} demonstrated that existing wire protocols carry redundant metadata and impose expensive per-tuple serialization, proposing protocol-level redesigns to mitigate the cost.
Federated query systems such as XDB~\cite{gavriilidis2023xdb}, Wayang~\cite{beedkar2023apache}, Dingo~\cite{giannakouris2150rethinking}, and MuSQLE~\cite{giannakouris2016musqle} optimize \emph{where} different parts of a query should be executed, yet they still rely on the wire protocol to external table transfer.
In all of these approaches, the database server remains the sole gatekeeper of its own storage. No matter how aggressively the client or middleware is optimized, every byte of data must still pass through the wire protocol.
 
We observe that this bottleneck is not inherent to the data itself, but to the \emph{access path}.
Database storage formats, including PostgreSQL's heap files, MySQL's InnoDB \texttt{.ibd} pages, are complex, but they are not opaque.
Their layouts are fully specified by a combination of source code, official documentation, and header files that have been stable across major releases.
A reader that can parse these files directly and emit columnar in-memory byte streams would remove the server from the critical path entirely, turning what was an engine-mediated query into a simple file scan.
 
The main barrier to building custom table readers is engineering effort: hand-writing a correct parser for even one database's on-disk format requires deep knowledge of page layouts, which might include MVCC visibility rules, type encodings, and TOAST or overflow-page handling.
Extending that parser to a second database doubles the effort, and keeping it current across versions adds a continuous maintenance burden.
We argue that recent advances in Large Language Models (LLMs) change this calculus.
An LLM can ingest the same documentation and source code that a human developer would consult and generate a format-specific reader in a single automated pass.
If the generated code fails to compile or produces incorrect output, an agentic feedback loop can diagnose the error and regenerate the code, reducing what was a weeks-long systems programming task to a pipeline that completes in minutes.
 
In this paper we explore the ability of LLMs to synthesize custom, high-performance table readers for a target database system that can read directly their files. First, we evaluate this idea by synthesizing two custom readers for Postgres and MySQL. We showcase that synthesizing custom table readers for already existing database system with LLMs is possible, and can achieve up to 27$\times$ speedups. Second, we present the vision of \textbf{Jailbreak}, an agentic approach that automates the synthesis of high-performance table readers that bypass the database engine entirely. Jailbreak-generated table readers ingest storage files directly and materialize table data as Apache Arrow columnar buffers.
Jailbreak's multi-agent LLM pipeline takes a table schema and database type as input, and aims to produce a verified, plug-in-compatible shared library implementing a \texttt{pg\_to\_arrow} (or \texttt{mysql\_to\_arrow}) C interface.
The pipeline comprises four specialized agents:
\begin{itemize}
    \item A Dataset Generator that creates test tables consisting of various attributes and data types.
    \item An Architect that produces a structured table reader format specification.
    \item A Coder that generates high-performance, C++ reader code.
    \item A QA Tester that verifies correctness against known ground truth using the target database.
\end{itemize}

All the above are connected by an iterative feedback loop that refines the generated code until all checks pass or a retry budget is exhausted.
 
The resulting reader produces Arrow \texttt{RecordBatch} buffers that are consumable in a zero-copy fashion by the majority of analytical query engines, including DuckDB, Apache Spark, DataFusion, PyArrow, and GPU-accelerated frameworks such as cuDF and Spark RAPIDS. Because the reader operates on storage files rather than through the database server, it eliminates the JDBC/ODBC protocol and query execution overheads from the data path. We validate correctness against JDBC/ODBC-based baselines using the full TPC-H benchmark and demonstrate speedups of up to ~5$\times$ for PostgreSQL and ~27$\times$ for MySQL across six analytic query engines. To the best of our knowledge, Jailbreak is the first system to leverage LLM-assisted code synthesis for database bypass. Our experimental evaluation showcase that this methodology is plausible and generalizable: any database system whose file format is recoverable from documentation or source code is a candidate for automated reader generation, opening a path toward breaking data lock-in without reverse engineering or vendor cooperation.
 
\smallskip
\noindent In summary, we make the following contributions:
\begin{itemize}
    \item We identify direct storage-file access as a principled alternative to wire-protocol-based data export, removing the database engine from the analytical data path.
    \item We propose a four-agent LLM pipeline for automated synthesis of correct, type-complete C++ readers for PostgreSQL heap files and MySQL InnoDB pages, and demonstrate its viability through LLM-assisted interactive development
    \item We produce Apache Arrow buffers that integrate zero-copy with six downstream engines, including GPU-accelerated frameworks.
\item We evaluate two synthesized readers on TPC-H, demonstrating end-to-end ETL speedups up to ~5.1$\times$ over PostgreSQL and up to ~27$\times$ over MySQL wire-protocol baselines, with full correctness validation across all 22 TPC-H queries.
\end{itemize}
\section{Overview}
\label{sec:overview}
\subsection{Jailbreak}
Jailbreak is an agentic approach for synthesizing high-performance table readers for database system bypass. Jailbreak exploits the two following insights. First, the storage layer details of a database system can be ingested by an LLM both from the source code and documentation. For most of the popular open-source database systems, this information is already included in the already existing pretrained LLMs. Second, the LLM can utilize this knowledge in order to automatically synthesize high-performance code that can read the database system storage files directly, by completely bypassing the database system. This approach offers several performance benefits. First, it eliminates the wire protocol serialization/deserialization overheads, since the external application or query engine can read the storage files directly. Second, it leaves the target database system intact, since query execution that normally happens through JDBC/ODBC and adds extra load to the database system is completely eliminated. Third, it gives more control to the higher application layer on \emph{how} to read and handle the data (e.g. partitioning schemes), since the synthesized reader works at the byte-level.

\subsection{Implementation}
We implement two shared libraries, one for Postgres (\texttt{pg\_arrow.so}) and one for MySQL (\texttt{mysql\_arrow.so}),
each compiled from a single self-contained C++17 translation unit
(approximately 800 lines each) with no external dependencies beyond the
standard library.
Both export a C-linkage function that reads a database storage file directly
from disk and returns a fully typed Apache Arrow \texttt{RecordBatch} via the
Arrow C Data Interface\footnote{\href{https://arrow.apache.org/docs/format/CDataInterface.html}{https://arrow.apache.org/docs/format/CDataInterface.html}} without any involvement of the database server, wire protocol or serialization. In the following sections we present the abstract Arrow C Data Interface, as well as the implementation details of two synthesized table readers, including Postgres and MySQL.

\subsection{Arrow C Data Interface}

The Arrow C Data Interface defines two plain C structs,
\texttt{ArrowSchema} and \texttt{ArrowArray}, that carry a complete
columnar \texttt{RecordBatch} across language and library boundaries without
copying the underlying buffers.
Each library inlines the two struct definitions verbatim, making the
compiled \texttt{.so} independent of any particular Arrow SDK version.
The consumer (Python, DuckDB, DataFusion, etc.) imports the batch with a
single call:

\begin{lstlisting}[language=Python, basicstyle=\small\ttfamily]
batch = pa.RecordBatch._import_from_c(
            ctypes.addressof(array), ctypes.addressof(schema))
\end{lstlisting}

\noindent
This zero-copy handoff is the key performance property: the column buffers
allocated by the C++ reader are owned by the \texttt{ArrowArray} through a
release callback and are freed only when the consumer releases the batch.

Table~\ref{tab:arrow-types} summarizes the type mapping used by both readers.
Fixed-width types (booleans, integers, floats, dates, timestamps) use
Arrow's two-buffer layout: a null bitmap and a packed data buffer.
Variable-width types (strings, binary) use three buffers: a null bitmap,
a 32-bit integer offset array, and a flat byte array for the character data.

\begin{table}[h]
\centering
\caption{Source type to Arrow format string mapping.}
\label{tab:arrow-types}
\small
\begin{tabular}{lll}
\hline
\textbf{Source type} & \textbf{Arrow format} & \textbf{Description} \\
\hline
\texttt{bool}              & \texttt{b}    & 1-bit boolean \\
\texttt{int2 / smallint}   & \texttt{s}    & int16 \\
\texttt{int4 / int}        & \texttt{i}    & int32 \\
\texttt{int8 / bigint}     & \texttt{l}    & int64 \\
\texttt{float4 / real}     & \texttt{f}    & float32 \\
\texttt{float8 / double}   & \texttt{g}    & float64 \\
\texttt{date}              & \texttt{tdD}  & date32 (Unix epoch) \\
\texttt{timestamp}         & \texttt{tsu:} & timestamp[\textmu s, UTC] \\
\texttt{numeric / decimal} & \texttt{g}    & float64 (decoded) \\
\texttt{bpchar / varchar / text} & \texttt{u} & UTF-8 string \\
\hline
\end{tabular}
\end{table}

\subsection{PostgreSQL Heap Reader}

\subsubsection{Page Layout}

PostgreSQL stores each table as one or more \emph{heap segment files} in
\texttt{\$PGDATA/base/\textit{dboid}/\textit{reloid}} (splitting at 1\,GB
boundaries with suffixes \texttt{.1}, \texttt{.2}, \ldots).
Each file is a sequence of 8\,KB pages.
A page begins with a 24-byte \texttt{PageHeaderData} that contains
\texttt{pd\_lower}, the byte offset of the end of the line pointer array;
\texttt{pd\_upper}, the start of tuple data; and \texttt{pd\_special}, the
start of any special-purpose area (unused for heap pages).
Following the header is an array of 4-byte \emph{item identifiers} (\texttt{ItemId}),
each encoding the page-relative offset, flags, and length of one tuple.
Tuples are stored in reverse order from the top of the page downward.

\subsubsection{MVCC Visibility}

Each heap tuple begins with a 23-byte \texttt{HeapTupleHeader}.
The implementation performs a lightweight MVCC visibility check before
decoding: a tuple is skipped if its \texttt{t\_xmax} (delete transaction ID) is
non-zero \emph{and} the \texttt{infomask} flags do not indicate that the deletion
was rolled back (\texttt{HEAP\_XMAX\_INVALID}) or was only a row-lock rather
than a real delete (\texttt{HEAP\_XMAX\_LOCK\_ONLY}).
This correctly excludes deleted and in-progress rows while retaining all
committed live tuples, mirroring the visibility semantics of a serializable
snapshot without requiring access to the transaction status tables.

\subsubsection{Tuple Decoding}

Each tuple's attribute data begins at byte offset \texttt{t\_hoff}, which
accounts for the header and any null bitmap.
Attributes are decoded sequentially.
Fixed-width types are aligned to their natural boundary (1, 2, 4, or 8 bytes)
and copied directly from the page.
Two epoch conversions are applied inline: PostgreSQL \texttt{date} stores days
since 2000-01-01, so $10{,}957$ days are added to obtain the Arrow
\texttt{date32} (days since Unix epoch); similarly, \texttt{timestamp} values
are shifted by $10{,}957 \times 86{,}400 \times 10^6$ microseconds.

Variable-length data (\texttt{bpchar}, \texttt{varchar}, \texttt{text}) uses
PostgreSQL's \emph{varlena} encoding.
When the low-order bit of the first byte is set the datum is in ``short''
form: the remaining 7 bits of that byte encode the total datum size (including
the 1-byte header), giving a maximum inline length of 127 bytes.
Otherwise the full 4-byte header is read and the payload length is
\texttt{(header >> 2) - 4}.
Trailing spaces are stripped from \texttt{bpchar} values before insertion into
the Arrow character buffer.

\texttt{numeric} (arbitrary-precision decimal) is decoded to \texttt{float64}.
PostgreSQL's internal numeric format stores digits in base~$10{,}000$
(\emph{NBASE} groups), with a 2- or 4-byte header encoding the sign, the
weight (most-significant NBASE group exponent), and the display scale;
the number of digit groups is inferred from the remaining payload length.
Both \emph{short} (2-byte header, weight stored as a 6-bit magnitude with a
sign-extension bit) and \emph{long} (4-byte header, explicit 16-bit signed
weight) formats are handled.
The decode accumulates the digit groups and adjusts by
$10{,}000^{\,\texttt{weight}-\texttt{ndigits}+1}$.

\subsubsection{I/O Strategy}

The reader buffers I/O in 2\,MB batches (256 pages per \texttt{fread} call)
to amortize system-call overhead.
Column buffers are pre-allocated assuming approximately 6.5\,million rows to
avoid repeated reallocation during the scan.
Row count is inferred from the first column's buffer length after the scan,
so the caller does not need to provide it in advance.

\subsubsection{BRIN-Pruned Scan}

The \texttt{pg\_to\_arrow\_brin} entry point accepts a path to the BRIN index
file and a predicate of the form \texttt{"col:type:min:max"}, where either
bound may be empty to indicate an unbounded range.
The BRIN file is loaded entirely into memory (typically a few hundred
kilobytes) and parsed as follows.
Page~0 is the \emph{meta page}: it holds the magic number
(\texttt{0xA8109CFA}), the \texttt{pagesPerRange} value, and the index of
the last revmap page.
Pages 1 through \texttt{lastRevmapPage} are \emph{revmap pages}, each storing
an array of 6-byte \texttt{ItemPointerData} entries at a fixed layout; the
capacity of a revmap page is $(\texttt{pd\_special} - 24) / 6$ rather than
being derived from \texttt{pd\_lower}, which PostgreSQL does not update on
revmap writes.

Each \texttt{ItemPointer} is followed to its \texttt{BrinTuple}.
A critical subtlety in the \texttt{bt\_info} field: the lower byte encodes
two bits per stored attribute, bit $2i$ indicates that attribute $i$ is
entirely null (no valid min/max), while bit $2i+1$ indicates the presence of
\emph{some} nulls but still valid extrema.
A page range is pruned only when its summary tuple has valid (non-all-null)
min and max values \emph{and} the stored
$[\texttt{brin\_min}, \texttt{brin\_max}]$ interval is disjoint from the
predicate range.
If either the min or max attribute is marked all-null, the BRIN summary is
treated as uninformative and the range is conservatively scanned.
Pruned page ranges are realized as \texttt{fseek} calls in
\texttt{do\_heap\_scan}, providing genuine I/O reduction rather than merely
parse-time filtering.

\subsection{MySQL InnoDB Reader}

\subsubsection{Page Layout}
MySQL InnoDB stores each table in a \texttt{.ibd} tablespace file composed of
16\,KB pages.
Each page begins with a 38-byte FIL header whose byte offsets 24--25 contain
the 2-byte big-endian page type; only \texttt{FIL\_PAGE\_INDEX} pages
(\texttt{0x45BF}) are processed.
Index pages carry a 36-byte \texttt{PAGE\_HEADER} followed by 20 bytes of
file-segment (FSEG) headers, after which the infimum and supremum system
records begin.
The high bit of \texttt{PAGE\_N\_HEAP} at offset~42 indicates the
\emph{COMPACT} row format.
\texttt{PAGE\_LEVEL} at offset~64 distinguishes B-tree leaf pages (level~0)
from interior nodes; only leaf pages contain row data.

\subsubsection{Record Traversal}

Within a leaf page, records form a singly linked list ordered by primary key.
The \emph{infimum pseudo-record} at page offset~99 serves as the list head;
its 5-byte record header stores a big-endian signed 16-bit \emph{next record}
offset at bytes~3--4.
Adding this offset to the current record's \emph{origin pointer} (the address
immediately following the header) gives the next record's origin.
Traversal terminates at the \emph{supremum} pseudo-record at offset~112 or
when the next-record field is zero.
Records flagged deleted (bit~5 of the header's info\_bits byte) and records
whose type field (the low 3 bits of the third header byte) is non-zero ---
that is, any record other than an ordinary leaf row, such as B-tree node
pointers or the system pseudo-records --- are skipped without decoding.

\subsubsection{Compact Row Format Prefix}

Before each record's origin, InnoDB stores two structures in reverse column
order that are essential for correct decoding.
First, the \emph{variable-length prefix array}: one or two bytes per
variable-length column (one byte if the maximum byte length $\le 255$,
two bytes otherwise), stored so the entry for the \emph{first} variable-length
column is closest to the origin.
Second, the \emph{null bitmap}: one bit per nullable column, also in reverse
order, stored immediately before the 5-byte record header.
Because the first variable-length column lies nearest the origin, the
implementation reads these lengths by starting just before the null bitmap
(adjacent to the record header) and decrementing its cursor backward through
memory for each successive column, rather than scanning forward.

\subsubsection{Type Decoding}

InnoDB stores all numeric types in \emph{big-endian} byte order with a sign
flip applied to enable byte-wise comparison.
For signed integers the sign bit is XOR'd with~1 (e.g., \texttt{0x80000000}
for \texttt{int4}), so that $-2^{31}$ maps to \texttt{0x00000000} and
$+2^{31}-1$ maps to \texttt{0xFFFFFFFF}.
Floating-point values use a two-case XOR: if the sign bit is set (the value is
positive in IEEE~754), only the sign bit is flipped; otherwise all bits are
inverted.
After de-mangling the bytes are reinterpreted natively via \texttt{memcpy}.

\texttt{DECIMAL(M,D)} is stored in MySQL's binary decimal format.
The integer and fractional parts are each split into groups of nine decimal
digits, where each full group occupies 4 bytes and a partial group occupies
1--4 bytes according to the lookup table
\texttt{DIG2BYTES}$[0..9] = [0,1,1,2,2,3,3,4,4,4]$.
For the integer part the partial group is the most-significant one and is
stored first; for the fractional part the partial group is the
least-significant one and is stored last.
The sign is encoded in the high bit of the first byte (1 = positive) and
removed before decoding; negative values additionally require flipping all
bytes.

\texttt{DATE} occupies 3 bytes with the sign encoded in the high bit of the
first byte (XOR \texttt{0x80}).
The remaining 23 bits pack year (14 bits), month (4 bits), and day (5 bits).
These are extracted and converted to a Julian Day Number, then adjusted to
the Unix epoch by subtracting the JDN of 1970-01-01 (2,440,588), yielding an
Arrow \texttt{date32} value.

InnoDB's \emph{hidden columns} --- the 6-byte transaction ID (\texttt{TRX\_ID})
and 7-byte rollback pointer (\texttt{ROLL\_PTR}) that separate primary-key
data from non-key data in the physical record --- are handled via a
\texttt{raw:13} skip entry in the column specification, which advances the
data pointer without emitting any output column.

\subsection{Shared Design Principles}

Both readers are intentionally structurally similar.
Column specifications are passed as a comma-separated string of
\texttt{"name:type"} pairs that is parsed at runtime, making the library
table-agnostic.
All heap or tablespace traversal, type decoding, and Arrow buffer construction
happens in a single sequential pass per file, minimizing memory traffic.
Buffers for fixed-width and variable-width columns are maintained as
\texttt{std::vector} containers and transferred to the Arrow structs by
pointer without copying.
The release callback registered on the top-level \texttt{ArrowArray}
recursively frees all child \texttt{ArrowArray}/\texttt{ArrowSchema}
objects and the owning \texttt{BatchData} allocation, so the caller is
responsible only for releasing the root struct.

\section{Agentic C++ Reader Generation}
\label{sec:agentic}
In order to automate the process of synthesizing a table reader, we present a multi-agent pipeline that takes a table name as input and outputs a verified, plug-in-compatible shared library (\texttt{pg\_arrow\_agent.so}) that implements the same \texttt{pg\_to\_arrow}
C interface as the hand-written reader described in Section~\ref{sec:overview}.

\subsection{Pipeline Overview}

\begin{figure}[h]
\centering
\begin{tikzpicture}[
    node distance=0.9cm,
    box/.style={rectangle, draw, rounded corners, minimum width=3.2cm,
                minimum height=0.8cm, align=center, font=\small},
    arrow/.style={->, thick}
]
\node[box] (dg)   {Dataset Generator};
\node[box, below=of dg]   (arch) {Architect};
\node[box, below=of arch] (eng)  {Software Engineer};
\node[box, below=of eng]  (qa)   {QA Tester};

\draw[arrow] (dg)   -- (arch);
\draw[arrow] (arch) -- (eng);
\draw[arrow] (eng)  -- (qa);
\draw[arrow] (qa.east) .. controls +(1.2,0) and +(1.2,0) ..
    node[right, font=\scriptsize]{fail} (eng.east);
\end{tikzpicture}
\caption{Agentic pipeline. The Coder--QA Tester loop repeats until
         the reader passes all correctness checks or the retry budget is
         exhausted.}
\label{fig:agentic-pipeline}
\end{figure}
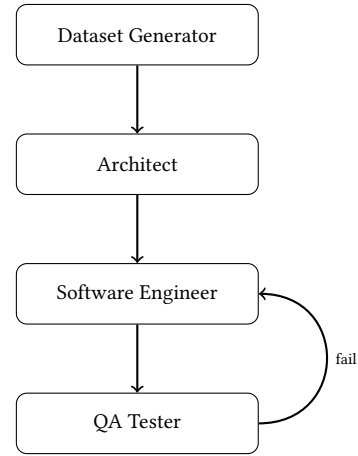

The pipeline consists of four specialized agents that execute in sequence, with an iterative feedback loop between the code-generation and verification stages (Figure~\ref{fig:agentic-pipeline}).

\paragraph{Dataset Generator.}
Rather than developing against a full benchmark table, the pipeline first creates a small synthetic test table ($\approx$50 rows) with the same schema as the target table of the database. However, the resulting reader can be used and it's compatible with any table in the database. Row values are generated deterministically, so the exact expected aggregates (min, max, sum for numeric columns; distinct count for string columns; min/max for dates) are known \emph{before} the reader runs. This makes correctness verification a simple arithmetic comparison rather than a probabilistic check.

\paragraph{Architect.}
Given the database type and table schema (retrieved via
\texttt{information\_schema}), the Architect agent produces a structured JSON
\emph{format spec} that encodes all information the Coder agent needs:
page-level constants (page size, header offsets, magic numbers), per-column
type encodings (endianness, decode formulae, varlena header layout), Arrow
format strings for each column, MVCC visibility rules, and a curated list of
known non-obvious bugs (\texttt{known\_gotchas}).
The agent's system prompt is grounded with a project knowledge base
(\texttt{knowledge/pg\_format.md}) covering the PostgreSQL heap page layout
and the Arrow C Data Interface buffer requirements.

\paragraph{Coder.}
The Coder agent receives the format spec and generates a complete,
self-contained C++ source file.
The output must compile with \texttt{g++ -O3 -std=c++17 -shared -fPIC} and
export the following C function:

\begin{lstlisting}[language=C, basicstyle=\small\ttfamily]
int pg_to_arrow(const char* heap_path, const char* col_spec,
                ArrowSchema* out_schema, ArrowArray* out_array,
                char* errmsg, int errmsg_len);
\end{lstlisting}

The agent is required to (i)~parse \texttt{col\_spec} at runtime rather than
hardcoding column names or types, and (ii)~implement all twelve standard
PostgreSQL types (\texttt{bool}, \texttt{int2/4/8}, \texttt{float4/8},
\texttt{numeric}, \texttt{date}, \texttt{timestamp}, \texttt{bpchar},
\texttt{varchar}, \texttt{text}) regardless of which subset appears in the
target table.
The system prompt explicitly forbids prose output; the first character of the
response must be \texttt{\#} or \texttt{/} (a preprocessor directive or
comment), and a post-processing step strips any residual natural-language text
before the source is written to disk.

On retry attempts the agent additionally receives the full compiler output, the
runtime output, and the complete history of prior QA diagnoses.
If the same error type recurs across two or more consecutive attempts, the
prompt instructs the agent to discard the current subsystem and rewrite it
from scratch using a different strategy.
A hash of each generated source is tracked to detect stuck loops in which the
model produces identical code despite receiving failure feedback; in such cases
an explicit rewrite hint is injected into the next prompt.

\paragraph{QA Tester.}
The QA Tester agent compiles the generated source, loads the resulting shared
library via \texttt{ctypes}, runs it against the synthetic heap file, and
imports the Arrow output as a PyArrow \texttt{RecordBatch}.
Correctness is checked deterministically: row count and per-column statistics
are compared against the Python-computed ground truth with a relative tolerance
of $10^{-3}$ for floating-point values.
Only if a mismatch is detected does the agent invoke an LLM to interpret the
failure; the LLM is shown the raw compiler output, the runtime output, the
ground truth, and the history of already-fixed bugs, and is asked to produce a
structured JSON diagnosis with categorized new bugs and confirmed fixes.
This design keeps LLM calls on the critical path to a minimum: a correctly
functioning reader requires no LLM invocation in the QA step at all.

\subsection{Retry Loop and Checkpointing}

The Coder and QA Tester agents alternate for up to
$N_{\max}$ attempts (default $N_{\max}=10$).
After every failed attempt the full pipeline state, including format spec, heap path, column specification, ground truth, complete feedback history, and the last generated source, is serialized to \texttt{outputs/\textit{table}\_checkpoint.json}. If the retry budget is exhausted without a passing reader, a subsequent invocation with \texttt{--resume} reloads this checkpoint, skips the Dataset Generator and Architect stages, and starts a fresh $N_{\max}$-attempt loop with the accumulated feedback history intact.

\subsection{Compatibility and Deployment}

The generated library is designed to be a drop-in replacement for the
hand-written reader. It exports the identical \texttt{pg\_to\_arrow} symbol with the same function signature, the same inline \texttt{ArrowSchema} and \texttt{ArrowArray} struct layout, and the same zero-copy Arrow C Data Interface contract. Upon a successful run the pipeline automatically copies the winning shared library, e.g. \texttt{pg\_arrow\_agent.so}, in the project root, making it immediately usable by all existing benchmarks.

\subsection{Multi-Provider Support}

All agents are invoked through LiteLLM, which provides a unified interface
over OpenAI, Anthropic, AWS Bedrock, and local Ollama models.
Each of the three LLM-backed agents (Architect, Coder, QA Tester)
can be assigned an independent model via command-line flags, enabling
heterogeneous configurations such as using a reasoning-optimized model for
code generation while routing cheaper calls to a smaller model for diagnosis.
All agents run at temperature~0 to maximize determinism.

\section{Experimental Evaluation}
We present an early experimental evaluation of Jailbreak, in which we benchmark an end-to-end pipeline that extracts a table from a database system and exports it as a local Parquet file. We use Postgres and MySQL as the target database system and multiple query engines, including DuckDB, DataFusion, PyArrow, Spark SQL, as well as GPU-accelerated query engines including cuDF and Spark RAPIDS.

\subsection{Setup}
All of our experiments were run on a g4dn.xlarge AWS instance with 4 vCPUs, 16 GiB of memory, 1 NVIDIA T4 GPU, and 125 GB of local NVMe SSD storage. We used TPC-H 1GB as our main benchmark, and we run an end-to-end extraction the \emph{lineitem} table from the target database to a local Parquet file. All LLM agents are invoked via LiteLLM. Experiments used Sonnet 4.6.

\subsection{ETL Benchmark: Database to Parquet}
\label{sec:etl}

We evaluate end-to-end ETL performance by reading the TPC-H \texttt{lineitem} table (6,001,215 rows, $\approx$880\,MB on disk) and writing the result to a Parquet file. Each method is run three times in an isolated subprocess; we report the best wall-clock time. We compare two data-access strategies across five analytics engines:

\begin{itemize}
  \item \textbf{Jailbreak} — the heap file is read directly via
        \texttt{pg\_to\_arrow} / \texttt{mysql\_to\_arrow}, producing an
        Apache Arrow \texttt{RecordBatch} that is handed off zero-copy to the
        downstream engine.
  \item \textbf{Wire Protocol} — the engine reads from the database through
        the standard client protocol (PostgreSQL wire protocol or MySQL
        JDBC/connector).
\end{itemize}

\subsection{PostgreSQL}

Figure~\ref{fig:etl-pg-cpp-vs-wire} shows results for PostgreSQL. Jailbreak consistently outperforms the wire protocol across all engines. DuckDB is the fastest overall at \textbf{5.46\,s} via Jailbreak, compared to 9.58\,s over wire (1.76$\times$ speedup). The largest gains appear with PyArrow and DataFusion, where the wire protocol incurs 23–24\,s due to server-side serialization, while Jailbreak reads the same data in under 7\,s (3.5$\times$ speedup). cuDF achieves the best Jailbreak time at \textbf{3.98\,s}, benefiting from GPU-accelerated Parquet encoding after an Arrow handoff. Spark and Spark RAPIDS are the slowest engines in both modes (32–55\,s), reflecting JVM startup and shuffle overhead rather than I/O cost.

\subsection{MySQL}
Figure~\ref{fig:etl-mysql-cpp-vs-wire} shows results for MySQL. The Jailbreak times are broadly similar to PostgreSQL (5.3–38.7\,s), confirming that the bottleneck is the downstream engine rather than the source database. The wire protocol results, however, diverge dramatically: PyArrow via JDBC takes \textbf{228.6\,s}, 26.9$\times$ slower than the 8.5\,s Jailbreak path, and this dominates the chart visually. DuckDB wire takes 44.2\,s vs.\ 6.8\,s (6.5$\times$). The disparity is larger than in PostgreSQL because MySQL's JDBC connector serializes each row individually, whereas the C++ reader decodes raw InnoDB pages at memory bandwidth. Only DataFusion wire (19.5\,s) and cuDF wire (16.7\,s) are closer to their C++ counterparts, as both use the ConnectorX driver which batches rows more aggressively.

\subsection{Speedup Summary}

Figure~\ref{fig:etl-speedup} summarizes the speedup of Jailbreak over the wire protocol for both databases. For PostgreSQL the speedup ranges from 1.3$\times$ (Spark RAPIDS) to 5.1$\times$ (cuDF), reflecting the relatively efficient PostgreSQL wire protocol. For MySQL the gains are larger across the board, peaking at \textbf{26.9$\times$} for PyArrow and 6.5$\times$ for DuckDB. The results demonstrate that bypassing the database server is most impactful when the wire protocol is the primary bottleneck, as is the case for row-oriented connectors such as MySQL JDBC, and remains beneficial even for well-optimized protocols such as PostgreSQL's binary COPY format.

%

\definecolor{cppblue}{HTML}{2166AC}
\definecolor{wiregray}{HTML}{92C5DE}
\definecolor{pgblue}{HTML}{2166AC}
\definecolor{mysqlred}{HTML}{D6604D}

\pgfplotsset{
  etl bar/.style={
    ybar,
    bar width=0.55cm,
    width=\textwidth,
    height=5.5cm,
    enlarge x limits=0.08,
    symbolic x coords={DuckDB,DataFusion,PyArrow,cuDF,Spark,{Spark RAPIDS}},
    xtick=data,
    xticklabel style={font=\small},
    ylabel style={font=\small},
    ytick align=inside,
    minor ytick={},
    grid=major,
    grid style={dashed, gray!25},
    axis line style={gray!60},
    tick style={gray!60},
    legend style={
      at={(0.01,0.99)}, anchor=north west,
      font=\small, draw=none, fill=none,
      /tikz/every even column/.append style={column sep=0.4em},
    },
    nodes near coords,
    nodes near coords style={font=\scriptsize, inner sep=1pt},
    nodes near coords align={above},
    every node near coord/.append style={
      /pgf/number format/.cd, fixed, precision=1,
    },
  }
}

\begin{figure*}[t]
\centering
\begin{tikzpicture}
\begin{axis}[
  etl bar,
  ymin=0,
  ylabel={Best time (s)},
  title style={font=\small\bfseries},
  title={PostgreSQL $\to$ Parquet \quad (TPC-H SF=1, 6\,M rows)},
  nodes near coords={\pgfmathprintnumber[fixed,precision=1]{\pgfplotspointmeta}s},
]
\addplot[fill=cppblue, draw=cppblue!70!black] coordinates {
  (DuckDB,5.458) (DataFusion,6.885) (PyArrow,6.592)
  (cuDF,3.976) (Spark,32.213) ({Spark RAPIDS},35.181)
};
\addplot[
  fill=wiregray, draw=wiregray!70!black,
  postaction={pattern=north east lines, pattern color=white!40!wiregray},
] coordinates {
  (DuckDB,9.583) (DataFusion,23.788) (PyArrow,23.549)
  (cuDF,20.077) (Spark,54.911) ({Spark RAPIDS},47.340)
};
\legend{Jailbreak (direct heap), Wire Protocol}
\end{axis}
\end{tikzpicture}
\caption{PostgreSQL ETL benchmark: time to read \texttt{lineitem} and write
  Parquet using Jailbreak (direct heap access) vs.\ wire protocol, per
  analytics engine. Lower is better. Best of 3 runs.}
\label{fig:etl-pg-cpp-vs-wire}
\end{figure*}
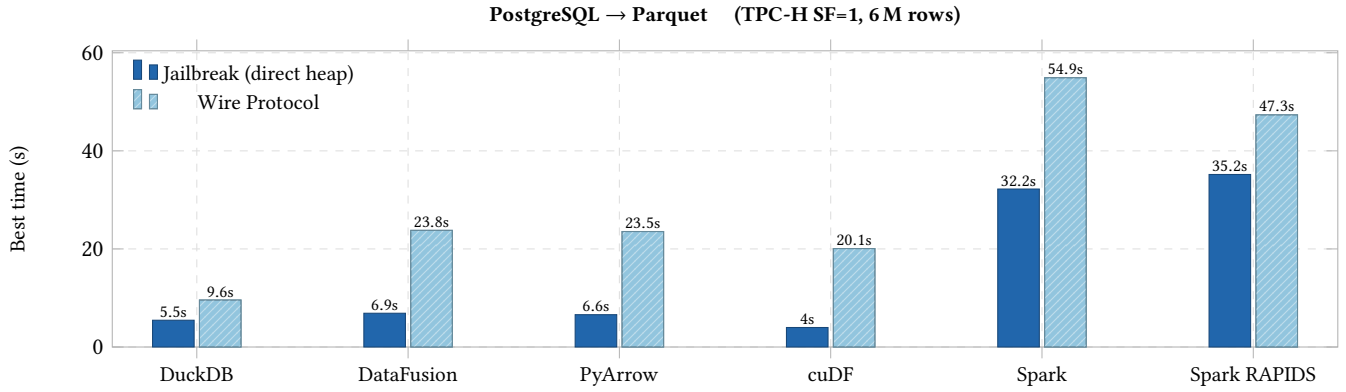

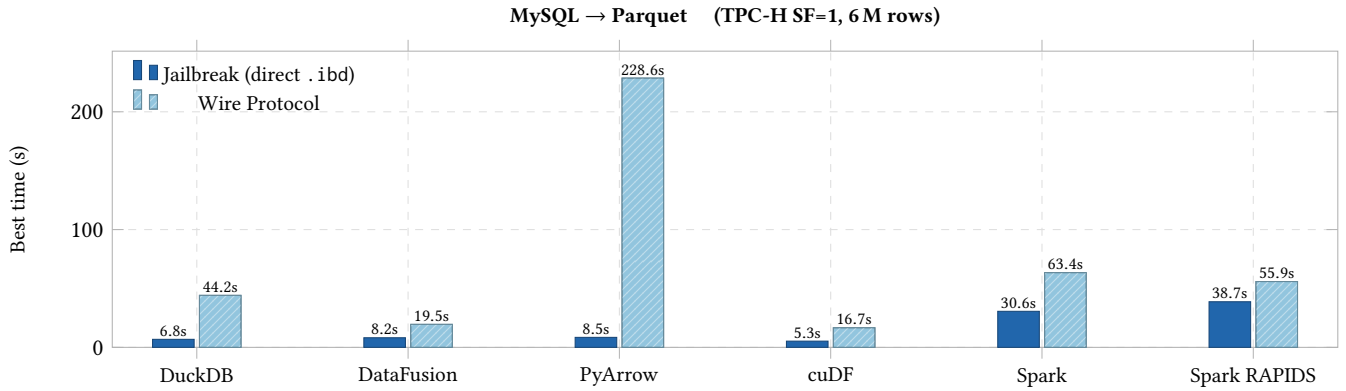
\begin{figure*}[t]
\centering
\begin{tikzpicture}
\begin{axis}[
  etl bar,
  ymin=0,
  ylabel={Best time (s)},
  title style={font=\small\bfseries},
  title={MySQL $\to$ Parquet \quad (TPC-H SF=1, 6\,M rows)},
  nodes near coords={\pgfmathprintnumber[fixed,precision=1]{\pgfplotspointmeta}s},
]
\addplot[fill=cppblue, draw=cppblue!70!black] coordinates {
  (DuckDB,6.810) (DataFusion,8.166) (PyArrow,8.479)
  (cuDF,5.277) (Spark,30.602) ({Spark RAPIDS},38.733)
};
\addplot[
  fill=wiregray, draw=wiregray!70!black,
  postaction={pattern=north east lines, pattern color=white!40!wiregray},
] coordinates {
  (DuckDB,44.160) (DataFusion,19.534) (PyArrow,228.604)
  (cuDF,16.661) (Spark,63.447) ({Spark RAPIDS},55.884)
};
\legend{Jailbreak (direct \texttt{.ibd}), Wire Protocol}
\end{axis}
\end{tikzpicture}
\caption{MySQL ETL benchmark: direct \texttt{.ibd} heap access vs.\ wire
  protocol per engine. PyArrow wire protocol takes 228\,s vs.\ 8.5\,s via
  the Jailbreak (26.9$\times$ speedup), dominating the y-axis.
  Best of 3 runs.}
\label{fig:etl-mysql-cpp-vs-wire}
\end{figure*}

\begin{figure*}[t]
\centering
\begin{tikzpicture}
\begin{axis}[
  etl bar,
  ymin=0,
  ylabel={Speedup ($\times$)},
  title style={font=\small\bfseries},
  title={Speedup of Jailbreak over Wire Protocol},
  nodes near coords={\pgfmathprintnumber[fixed,precision=1]{\pgfplotspointmeta}$\times$},
  extra y ticks={1},
  extra y tick style={grid=major, grid style={thick, red!60}},
  extra y tick labels={},
]
\addplot[fill=pgblue, draw=pgblue!70!black] coordinates {
  (DuckDB,1.756) (DataFusion,3.455) (PyArrow,3.573)
  (cuDF,5.050) (Spark,1.705) ({Spark RAPIDS},1.346)
};
\addplot[
  fill=mysqlred, draw=mysqlred!70!black,
  postaction={pattern=north east lines, pattern color=white!40!mysqlred},
] coordinates {
  (DuckDB,6.484) (DataFusion,2.392) (PyArrow,26.962)
  (cuDF,3.157) (Spark,2.073) ({Spark RAPIDS},1.443)
};
\legend{PostgreSQL, MySQL}
\end{axis}
\end{tikzpicture}
\caption{Speedup of Jailbreak direct heap access over wire protocol per engine.
  MySQL PyArrow achieves 26.9$\times$ because the MySQL wire protocol
  (JDBC/connector) is substantially slower than PostgreSQL's.}
\label{fig:etl-speedup}
\end{figure*}
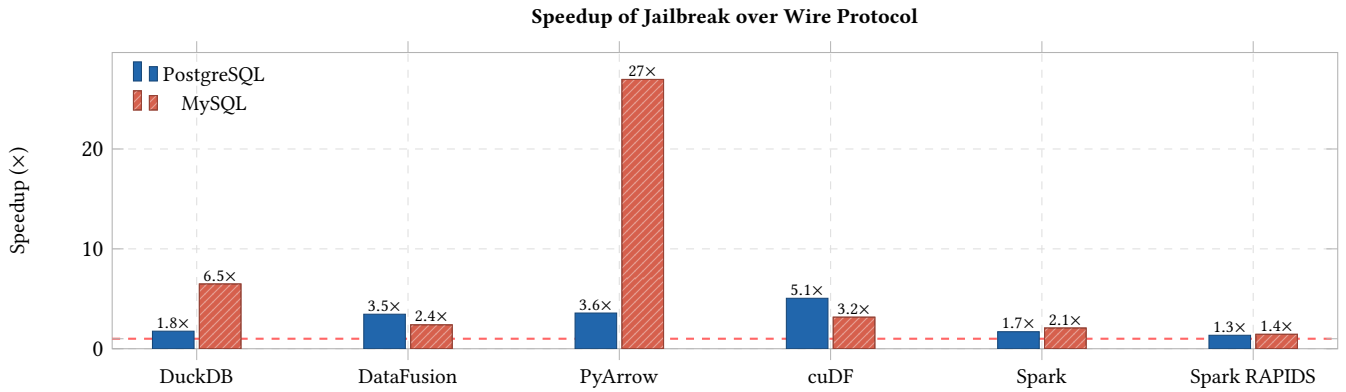

\section{Related Work}
\label{sec:related}

\noindent\textbf{Wire protocol overhead and data loading.}
The inefficiency of engine-mediated data access is well established.
Raasveldt and M\"{u}hleisen~\cite{raasveldt2017don} demonstrated that
existing wire protocols impose redundant metadata and expensive per-tuple
serialization costs, arguing for protocol-level redesigns.
ConnectorX~\cite{wang2022connectorx} identified client-side
deserialization as the dominant bottleneck in database-to-dataframe
transfers, accounting for over 85\% of wall-clock time in typical
\texttt{read\_sql} calls, and proposed parallel partitioned reads as
a mitigation. More recently, Karpathiotakis et al.~\cite{karpathiotakis2017no}
examined cross-system data movement in heterogeneous ecosystems,
and Palkar et al.~\cite{gavriilidis2025fast} explored fast, scalable
data transfer across decentralized data systems. These approaches
optimize the transfer path but retain the database server as the
gatekeeper of storage; Jailbreak eliminates the server from the
data path entirely.
\\
\textbf{Federated and multi-engine query processing.}
Federated systems such as XDB~\cite{gavriilidis2023xdb},
Wayang~\cite{beedkar2023apache}, Dingo~\cite{giannakouris2150rethinking},
and MuSQLE~\cite{giannakouris2016musqle} push query fragments to
heterogeneous engines to reduce data movement and improve
performance. These systems optimize \emph{where} computation runs,
but still rely on wire protocols for inter-engine data transport.
Jailbreak is complementary: it targets the data \emph{ingestion}
layer, producing Arrow buffers that federated engines can consume
directly without a wire-protocol hop.
\\
\textbf{Analytical engines and columnar formats.}
The emergence of embeddable analytical engines such as
DuckDB~\cite{raasveldt2019duckdb}, DataFusion~\cite{datafusion},
Velox~\cite{pedreira2022velox}, Spark SQL~\cite{zaharia2016apache},
and GPU-accelerated frameworks such as cuDF~\cite{cudf} has raised
expectations for zero-copy, columnar data delivery. Apache
Arrow~\cite{arrow} has become the de facto in-memory interchange
format for this ecosystem. Jailbreak is designed around this
standard: the generated readers emit Arrow RecordBatches via the
Arrow C Data Interface, making them immediately consumable by
any Arrow-native engine without format conversion.
\\
\textbf{LLM-assisted code synthesis.}
To the best of our knowledge, Jailbreak is the first system to
apply LLM-assisted code synthesis to database storage decoding.
Concurrent work explores LLM-driven synthesis more broadly in the
database systems context: GenDB~\cite{lao2026gendb} proposes a
vision of synthesizing entire query processing pipelines rather
than engineering them, and Wehrstein et al.~\cite{wehrstein2026bespoke}
demonstrate workload-specific OLAP engine synthesis tailored to
individual query patterns. These works share Jailbreak's core
premise that LLMs can generate correct, high-performance systems
code from specifications, but target query execution rather than
storage decoding. Jailbreak occupies a complementary niche:
the storage ingestion layer, where a well-defined binary format
and a deterministic correctness oracle enable automated
verification and iterative refinement.
\section{Conclusions}
\label{sec:conclusions}
We have presented \textbf{Jailbreak}, a system and methodology for
breaking database lock-in by synthesizing high-performance storage
readers that bypass the wire protocol entirely. Our key insight is
that database file formats, while complex, are fully recoverable from
source code and documentation, artifacts that LLMs can ingest to
generate format-specific readers without human-authored parsing logic.

We evaluated two LLM-assisted readers for PostgreSQL and MySQL,
demonstrating end-to-end ETL speedups of up to 5.1$\times$ and
27$\times$ respectively over wire-protocol baselines across six
analytical engines, with full correctness validation on all 22 TPC-H
queries. These results suggest that LLM-assisted storage layer
synthesis is a viable and generalizable methodology: any database
system whose file format is recoverable from documentation or source
code is a candidate for automated reader generation.

Beyond the readers evaluated here, we have implemented the
four-agent pipeline described in Section~\ref{sec:agentic} and are
actively working toward a fully automated, push-button workflow.
Our immediate next steps include a systematic evaluation of the
pipeline's convergence behavior, characterizing retry distributions,
success rates, and generation latency across a broader set of schemas
and database versions, as well as an ablation study assessing the
contribution of each agent to final reader correctness. We are also
exploring extension to additional database systems, including SQLite
and MongoDB, where file format specifications are similarly
recoverable from documentation and source code. We view Jailbreak
as a first step toward a broader vision in which data lock-in across
heterogeneous database systems can be broken without reverse
engineering or vendor cooperation.

\bibliographystyle{ACM-Reference-Format}
\bibliography{sample}

@article{wehrstein2026bespoke,
  title={Bespoke OLAP: Synthesizing Workload-Specific One-size-fits-one Database Engines},
  author={Wehrstein, Johannes and Eckmann, Timo and Jasny, Matthias and Binnig, Carsten},
  journal={arXiv preprint arXiv:2603.02001},
  year={2026}
}

@article{lao2026gendb,
  title={GenDB: The Next Generation of Query Processing--Synthesized, Not Engineered},
  author={Lao, Jiale and Trummer, Immanuel},
  journal={arXiv preprint arXiv:2603.02081},
  year={2026}
}

@article{gavriilidis2025fast,
  title={Fast and Scalable Data Transfer Across Data Systems},
  author={Gavriilidis, Haralampos and Beedkar, Kaustubh and Boehm, Matthias and Markl, Volker},
  journal={Proceedings of the ACM on Management of Data},
  volume={3},
  number={3},
  pages={1--28},
  year={2025},
  publisher={ACM New York, NY, USA}
}

@article{pedreira2022velox,
  title={Velox: Meta's Unified Execution Engine.},
  author={Pedreira, Pedro and Erling, Orri and Basmanova, Maria and Wilfong, Kevin and Sakka, Laith and Pai, Krishna and He, Wei and Chattopadhyay, Biswapesh},
  journal={Proc. VLDB Endow.},
  volume={15},
  number={12},
  pages={3372--3384},
  year={2022}
}

@inproceedings{raasveldt2019duckdb,
  title={Duckdb: an embeddable analytical database},
  author={Raasveldt, Mark and M{\"u}hleisen, Hannes},
  booktitle={Proceedings of the 2019 international conference on management of data},
  pages={1981--1984},
  year={2019}
}

@article{zaharia2016apache,
  title={Apache spark: a unified engine for big data processing},
  author={Zaharia, Matei and Xin, Reynold S and Wendell, Patrick and Das, Tathagata and Armbrust, Michael and Dave, Ankur and Meng, Xiangrui and Rosen, Josh and Venkataraman, Shivaram and Franklin, Michael J and others},
  journal={Communications of the ACM},
  volume={59},
  number={11},
  pages={56--65},
  year={2016},
  publisher={Acm New York, NY, USA}
}

@article{wang2022connectorx,
  title={ConnectorX: accelerating data loading from databases to dataframes},
  author={Wang, Xiaoying and Wu, Weiyuan and Wu, Jinze and Chen, Yizhou and Zrymiak, Nick and Qu, Changbo and Flokas, Lampros and Chow, George and Wang, Jiannan and Wang, Tianzheng and others},
  journal={Proceedings of the VLDB Endowment},
  volume={15},
  number={11},
  pages={2994--3003},
  year={2022},
  publisher={VLDB Endowment}
}

@article{raasveldt2017don,
  title={Don't hold my data hostage: a case for client protocol redesign},
  author={Raasveldt, Mark and M{\"u}hleisen, Hannes},
  journal={Proceedings of the VLDB Endowment},
  volume={10},
  number={10},
  pages={1022--1033},
  year={2017},
  publisher={VLDB Endowment}
}

@inproceedings{giannakouris2016musqle,
  title={MuSQLE: Distributed SQL query execution over multiple engine environments},
  author={Giannakouris, Victor and Papailiou, Nikolaos and Tsoumakos, Dimitrios and Koziris, Nectarios},
  booktitle={2016 IEEE International Conference on Big Data (Big Data)},
  pages={452--461},
  year={2016},
  organization={IEEE}
}

@article{giannakouris2150rethinking,
  title={Rethinking Pluggable Federated Query Optimization: From Laptops to Data Warehouses},
  author={Giannakouris, Victor and Trummer, Immanuel},
  journal={Proceedings of the VLDB Endowment. ISSN},
  volume={2150},
  pages={8097}
}

@article{gavriilidis2023xdb,
  title={XDB in Action: Decentralized Cross-Database Query Processing for Black-Box DBMSes.},
  author={Gavriilidis, Haralampos and Rose, Leonhard and Ziegler, Joel and Beedkar, Kaustubh and Quian{\'e}-Ruiz, Jorge-Arnulfo and Markl, Volker},
  journal={Proc. VLDB Endow.},
  volume={16},
  number={12},
  pages={4078--4081},
  year={2023}
}

@article{beedkar2023apache,
  title={Apache wayang: A unified data analytics framework},
  author={Beedkar, Kaustubh and Contreras-Rojas, Bertty and Gavriilidis, Haralampos and Kaoudi, Zoi and Markl, Volker and Pardo-Meza, Rodrigo and Quian{\'e}-Ruiz, Jorge-Arnulfo},
  journal={ACM SIGMOD Record},
  volume={52},
  number={3},
  pages={30--35},
  year={2023},
  publisher={ACM New York, NY, USA}
}

@inproceedings{karpathiotakis2017no,
  title={No data left behind: real-time insights from a complex data ecosystem},
  author={Karpathiotakis, Manos and Floratou, Avrilia and {\"O}zcan, Fatma and Ailamaki, Anastasia},
  booktitle={Proceedings of the 2017 Symposium on Cloud Computing},
  pages={108--120},
  year={2017}
}

@misc{datafusion,
  title        = {Apache DataFusion},
  howpublished = {\url{https://github.com/apache/datafusion}},
}

@misc{arrow,
  title        = {Apache Arrow},
  howpublished = {\url{https://github.com/apache/arrow}},
}

@misc{cudf,
  title        = {cuDF},
  howpublished = {\url{https://github.com/rapidsai/cudf}},
}

\end{document}